\documentclass[preprint,12pt]{elsarticle}
\usepackage{amssymb}
\usepackage{amsmath}
\journal{Physica D}

\begin{document}

\begin{frontmatter}

\title{Synchronization of branching chain of dynamical systems: application to the logistic map}

\author[inst1,inst2]{Michele Baia\corref{cor1}}
\cortext[cor1]{Corresponding author}
\ead{michele.baia@unifi.it}
\author[inst1,inst2]{Franco Bagnoli}
\ead{franco.bagnoli@unifi.it}
\author[inst1]{Tommaso Matteuzzi}
\ead{tommaso.matteuzzi@unifi.it}
\author[inst3]{Arkady Pikovsky}
\ead{pikovsky@uni-potsdam.de}

\newcommand{\orcidauthorA}{0000-0002-6293-0305} 
\newcommand{\orcidauthorC}{0000-0001-9682-7122} 

\affiliation[inst1]{organization={Department of Physics and Astronomy and CSDC, University of Florence},
                    addressline={G. Sansone, 1}, 
                    city={Sesto Fiorentino},
                    postcode={50019},
                    country={Italy}}

\affiliation[inst2]{organization={INFN, Sect.Florence},
                    addressline={G. Sansone, 1}, 
                    city={Sesto Fiorentino},
                    postcode={50019},
                    country={Italy}}
        
\affiliation[inst3]{organization={Institute of Physics and Astronomy, University of Potsdam},
                    addressline={Karl-Liebknecht-Str. 24/25}, 
                    city={Potsdam-Golm},
                    postcode={14476},
                    country={Germany}}            

\begin{abstract} 
We investigate the synchronization dynamics in a chain of coupled chaotic maps organized in a single-parent family tree, whose properties can be captured considering each parent node connected to two children, one of which also serves as the parent for the subsequent node. Our analysis focuses on two distinct synchronization behaviors: parent-child synchronization, defined by the vanishing distance between successive nodes along the chain, and sibling synchronization, corresponding to the convergence of the states of two child nodes.
Our findings reveal significant differences in these two type of  synchronization mechanisms, which are closely associated with the probability distribution of the state of  parent node. Theoretical analysis and simulations with the logistic map support our findings.
We further investigate numerical aspects of the implementation corresponding to cases for which the simulated regimes differ from the theoretically predicted one due to computational finite accuracy. We perform a detailed study on how instabilities are numerically suppressed or amplified along the chain. 
In some cases, a properly adjusted computational scheme can solve this problem.

\end{abstract}

\begin{keyword}

Master-slave (parent-child) synchronization \sep child-child synchronization \sep branching chain \sep logistic map \sep machine precision synchronization \sep spatio-temporal chaotic maps \sep instability due to numerical precision errors 

\end{keyword}

\end{frontmatter}


\section{Introduction}
Complete synchronization of chaotic systems has been first studied for two identical coupled units, whose states can become equal if the coupling is strong enough to suppress the chaotic instability. In spatially extended systems, such as coupled map lattices~\cite{Kaneko1992}, synchronization implies spatial homogeneity across the system. For such systems, one typically considers a local symmetric spatial coupling, though asymmetric forms have also been explored~\cite{SynchronizationBook}.

A notable example of asymmetric coupling is unidirectional or parent-child (alternatively referred to as master-slave) coupling of identical systems. Here, part of the signal from the parent system is injected into the child~\cite{Pecora1990,Pecora2015}, but there is no action in the opposite direction. The synchronization threshold is related to the chaotic properties of the uncoupled system. If the coupling strength exceeds a certain threshold, the states of the two systems become identical, provided that their parameters are the same. In many setups of this type of coupling configurations, there exists a relationship between the maximum Lyapunov exponent of the parent (the unperturbed system) and the critical coupling strength needed to achieve synchronization.

Furthermore, the same parent signal can be directed to multiple child systems, enabling studies of mutual synchronization among these ``children''. Here, the parent signal can be seen as a common noise source, albeit correlated with the states of  children. Intriguingly, noise-induced synchronization has been widely studied~\cite{Pikovsky-84a,SynchronizationBook,Goldobin-Pikovsky-05b,Flandoli2016}, yielding unexpected findings, such as the potential of randomness to promote synchronization even for chaotic conditions~\cite{Maritan1994}.
Recent researches expanded these foundational theories, focusing on synchronization generalizations. Notable advances include the study of phase synchronization~\cite{Rosenblum1996} and generalized synchronization~\cite{rulkov1995generalized}, these phenomena occur even with parameter mismatches. 

It is possible to extend the master-slave synchronization approach to spatio-temporal chaos, for instance coupled map lattices~\cite{Jiang1998}, spatio-temporal chaotic systems~\cite{Grassberger1999}, and cellular automata~\cite{Bagnoli1999}. These efforts aim to link synchronization behaviors to universal classes such as directed percolation and the KPZ class~\cite{Baroni2001,Ahlers-Pikovsky-02,Droz2003}. Furthermore, parental-child synchronization plays an important role in data assimilation, facilitating parameter extraction from time-series data~\cite{Bagnoli2023}.

\begin{figure}[!htb]
    \centering
   \includegraphics[width=0.7\linewidth]{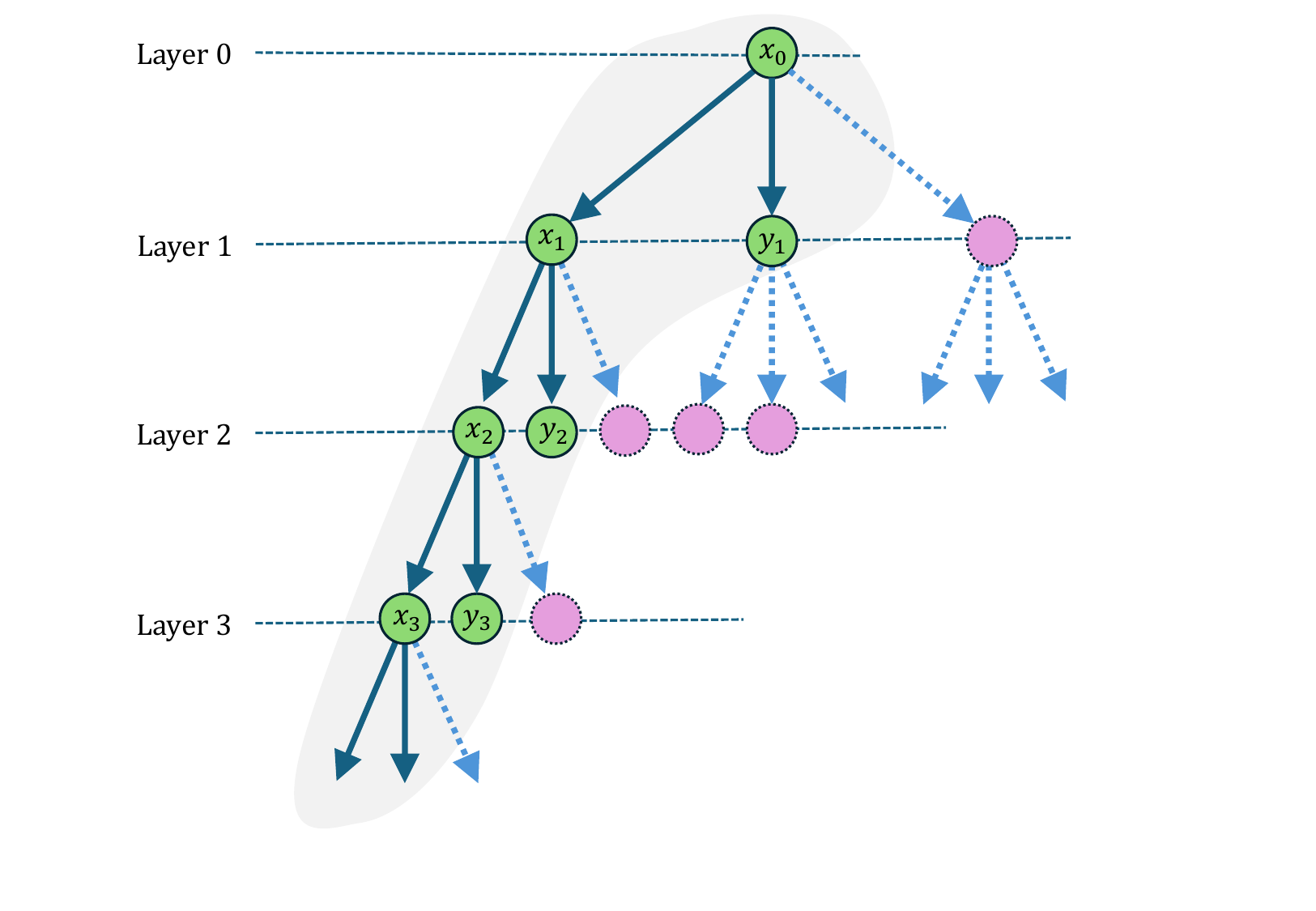}\\
    \caption{The selection of the branching chain. We start with a full Cayley tree (with levels denoted as ``layers'') with unidirectional coupling. For symmetry reasons, the behavior of all nodes in the same layer is the same, so we consider only two nodes (``siblings'') for each layer except for the top one (layer $0$).}
    \label{fig:TreeScheme}
\end{figure}

\begin{figure}[!htb]
    \centering
    \includegraphics[width=0.7\linewidth]{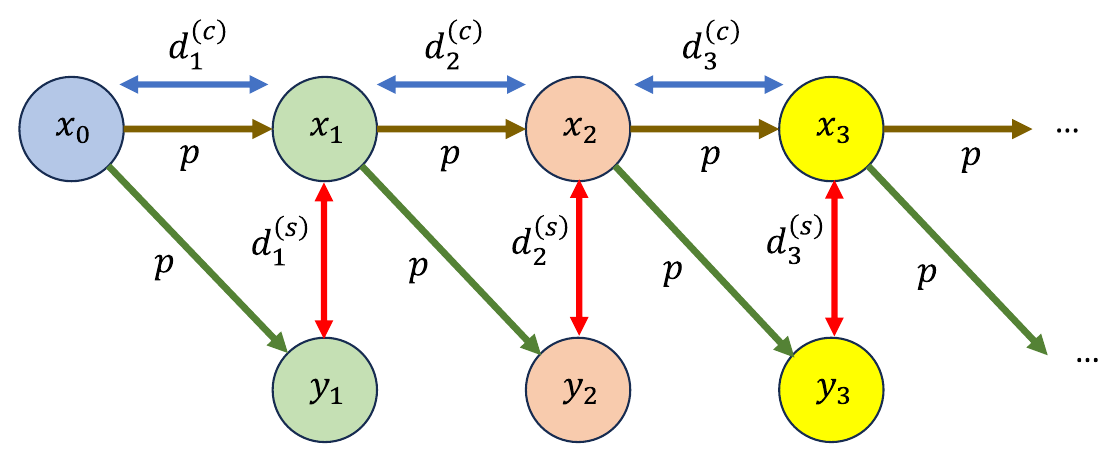}
    \caption{Branching chain scheme. The sub-index $i$ denotes layers. The arrow marked $p$ denotes the one-directional coupling (see Eq.~\eqref{eq:map}). The two-directional arrows denotes distances. Node labelled $x_0$ is the first parent. It has two ``children'', $x_1$ and $y_1$, and we measure the distance $d^{(c)}_1 = |x_0-x_1|$ between the parent and one of its children, and  $d^{(s)}_1 = |x_1-y_1|$ between the two ``siblings''. This scheme repeats itself at each layer $i$. }
    \label{fig:CouplingScheme}
\end{figure}

This work addresses the problem of parent-child synchronization in a unidirectional network structured as a Cayley tree (see Fig.~\ref{fig:TreeScheme}). The tree is organized in layers; the units at layer $i$ drive unidirectionally units of layer $i+1$, $i=0,1,2,\ldots$.  The symmetry of units within a layer allows us to assume equivalent behavior across nodes at the same layer, therefore we simplify the model by considering only two nodes (or “siblings”) per layer (excluding the root layer $i=0$).
This simplification reduces our study to a branching chain of chaotic maps, illustrated in Fig.~\ref{fig:CouplingScheme}. Our analysis focuses not only on the interaction between a parent and its child but also on the comparison of the states of siblings sharing a parent signal, assessing how siblings synchronize with each other and with their parent. Specifically, we study a chain of logistic maps as an illustrative case, which reveals distinctive synchronization behaviors.

In Section~\ref{sec:model}, we outline the model general structure, followed by a specific investigation of a logistic map chain in Section~\ref{sec:logistic}. 
This work is an extension of the preliminarily results presented in \cite{bagnoli2024ACRI}:
we will demonstrate that unexpected phenomena arise, particularly that siblings can achieve synchronization at much smaller coupling strengths than those required for parent-child synchronization. This behavior is linked to synchronization driven by common noise~\cite{Maritan1994, Flandoli2016,Baroni01}. Moreover, we highlight the impact of the computational method on simulating such chaotic coupled systems, noting that certain numerical schemes can introduce spurious dynamics, influenced by both the intrinsic chaotic nature of the system and the limitations of computational number representation.

\section{The branching chain model} \label{sec:model}
We consider one-dimensional maps $x(t+1)=f(x(t))$, where the variable $x\in[0,1]$ evolves in discrete time steps, indexed by the time index $t$. In the following, we shall indicate by $x \equiv x(t)$ and $x' \equiv x(t+1)$. 

Let us consider a branching chain scheme (Fig.~\ref{fig:CouplingScheme}) in which the first map is an autonomous ``pacemaker'' 
\[
  x_0' = f(x_0),
\]
and all other maps are driven by their ``parent map'' 
\begin{equation}\label{eq:map}
  \begin{split}
    x_i' &= (1-p) f(x_i) + p f(x_{i-1}),\\
    y_i' &= (1-p) f(y_i) + p f(x_{i-1}).\\
  \end{split}
\end{equation}
Here the parameter $p\in [0,1]$ defines the strength of the coupling and the lower index $i$ denotes the layer number.
This setup can be seen as the minimal sampling in a full Cayley tree, see Fig.~\ref{fig:TreeScheme}, where the first node $x_0$ is the root and for each subsequent layer $i>0$ only the two $x_i$ and $y_i$ nodes are selected.

For each layer, the two siblings $x_i$ and $y_i$ are guided by the same parent $x_{i-1}$ in the spirit of  the Pecora-Carrol scheme~\cite{Pecora1990,Pecora2015,Bagnoli2023} (when applied to maps). Such a setup is usual in the theory of generalized synchronization, where a ``replica'' of a driven element is introduced~\cite{rulkov1995generalized}.\\

This configuration allows us to define two types of synchronization:
\begin{itemize}
  \item the \textit{synchronization within a layer}, when $y_i(t)=x_i(t)$,
  \item the \textit{synchronization across the layers}, which means that $x_i(t)=x_{i-1}(t)=\ldots = x_0(t)$ for all $t$. 
\end{itemize}

The synchronization within a layer can be characterized by computing the average of the difference $d_i^{(s)}=|x_i-y_{i}|$:
\[
s_i= \langle d_i^{(s)} \rangle =\langle|x_i-y_{i}|\rangle = \frac{1}{T} \sum_{t=\tau}^{\tau+T} |x_i(t)-y_{i}(t)|,
\]
where $\tau$ is a transient interval.
When $s_i$ vanishes, it means that both ``siblings'' $x_i$ and $y_i$ follow the common drive $x_{i-1}$ in the same way.

Synchronization across the layers, on the other hand,  can be characterized by calculating the average in time  difference  $ d_i^{(c)}=|x_i-x_{i-1}|$:
\[
c_i=\langle d_i^{(c)} \rangle=\langle|x_i-x_{i-1}|\rangle=\frac{1}{T} \sum_{t=\tau}^{\tau+T} |x_i(t)-x_{i-1}(t)|,
\]
checking if this quantity vanishes. 

A transition to complete synchronization across all layers is expected at a critical value of the coupling parameter $p$, as $c_i > 0$ when $p=0$ and $c_i=0$ when $p=1$. The critical value for the synchronization across all the layers, independent of the index layer $i$, is designated as $p^{(c)}$. 

Synchronization across layers implies that the elements within each layer are also fully synchronized. However, as we will demonstrate, the reverse is not necessarily true: synchronization can occur within individual layers at coupling values below the threshold $p^{(c)}$. Furthermore, synchronization within certain layers can be achieved even as layers with higher indices $i$ remain desynchronize. 
This behavior for the logistic map is illustrated in Fig.~\ref{fig:distance_10}, which will be presented in details in the following.

Following standard approaches in complete synchronization theory, a criterion can be established by analyzing the behavior of small perturbations near the synchronized state~\cite{SynchronizationBook}. 
In order to do that, let us consider the synchronization of the first child $x_1$ with the parent $x_0$. Near the synchronization transition across the layers, $c_1\simeq 0$, we can write $x_1(t)=x_{0}(t)+\delta x_1(t)$ and, using Eq.~\eqref{eq:map}, we obtain:
\[
  \delta x_1' =  (1-p) \frac{\mathrm{d}{f}}{\mathrm{d}x}(x_{0})\delta x_1.
\]
Thus, the exponential growth rate of the perturbations in the correspondence of the complete synchrony is
\begin{equation}\label{eq:liapCteo}
\lambda_1^{(c)} = \ln(1-p) + \left\langle 
\ln\left|\frac{\mathrm{d}{f}}{\mathrm{d}x}(x_0)\right|\right\rangle= \ln(1-p) + \lambda_{0},
\end{equation}
where $\lambda_0=\left\langle 
\ln\left|\frac{\mathrm{d}{f}}{\mathrm{d}x}(x_0)\right|\right\rangle$ is the Lyapunov exponent of the driving map at level $0$. From the above expression, the theoretical estimate of the critical value of the coupling can be easily obtained:
\begin{equation}\label{eq:pic}
p^{(c)} = 1-\exp(-\lambda_0).
\end{equation}

Assuming that the parent-children synchronization starts from the top layers (first $x_1$ with $x_0$, then $x_2$ with $x_1$, etc.), this relation is valid for the first non-synchronized layer, and thus in principle all layers should experience synchronization transition at the same value of $p^{(c)}$.

We can repeat the analogous calculations for the siblings, computing the evolution of the divergence $\delta y_i$ of the trajectory as a function of the dynamic variable $x$ of the first sibling
\[
  \delta y_i' = y'_i-x'_i =  (1-p) \left(\frac{\mathrm{d}{f}}{\mathrm{d}x}(x_i)\right)(y_i-x_i) = (1-p) \left(\frac{\mathrm{d}{f}}{\mathrm{d}x}(x_i)\right)\delta y_i.
\]
The exponential growth rate of the perturbation can be written as:
\begin{equation}\label{eq:lambdab}
\lambda^{(s)}_i = \ln(1-p) + \left\langle \ln\left|\frac{\mathrm{d}{f}}{\mathrm{d}x}(x_{i})\right|\right\rangle= \ln(1-p) + \lambda_{i},
\end{equation}
and, therefore
\begin{equation}\label{eq:pis}
p_i^{(s)} = 1-\exp(-\lambda_i).
\end{equation}
Note that, since $x_i$ and $x_0$ are not synchronized in general, the quantity $\lambda_i=\left\langle \ln\left|\frac{\mathrm{d}{f}}{\mathrm{d}x}(x_{i})\right|\right\rangle$ defined above is not the Lyapunov exponent of the uncoupled map, but depends on the local dynamics of $x_i$.
This value can be calculated from the probability distribution $P_i(x)$ of the  variable $x_i$,
\begin{equation}\label{eq:lyapprob}
    \lambda_i = \int_0^1 \ln\left(\left|\frac{\mathrm{d}f}{\mathrm{d}x}(x)\right|\right) P_i(x)\mathrm{d}x.
\end{equation}

\section{Chain of logistic maps}\label{sec:logistic}
\begin{figure}[!t] 
    \centering
    \includegraphics[width=0.7\textwidth]{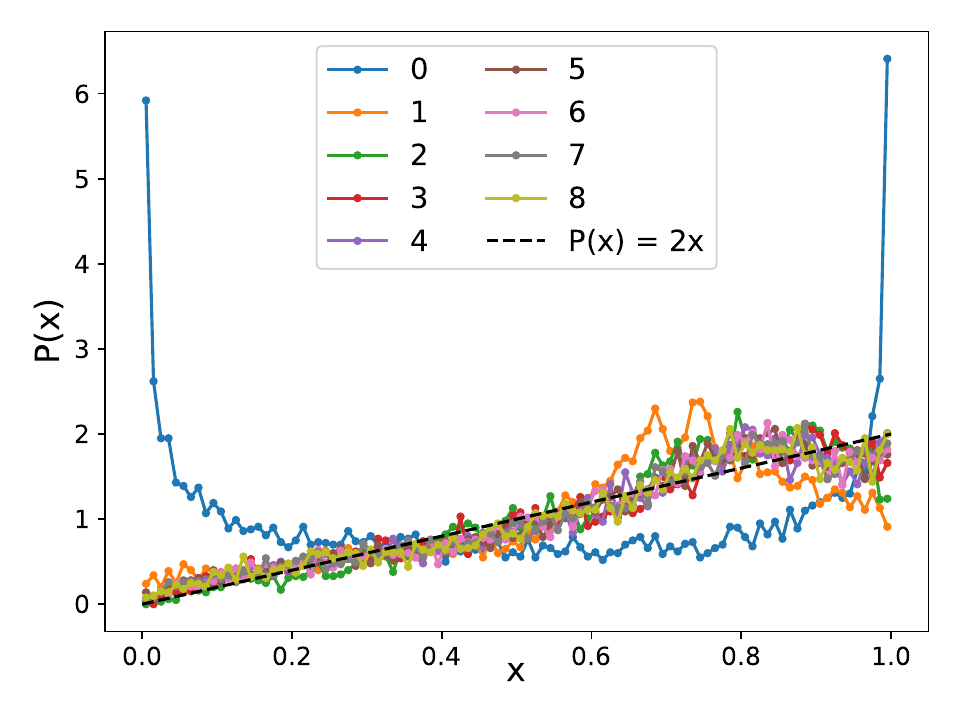}  \\
    \caption{Probability distribution $P_i(x)$ of maps along the chain for $p=0.32$, and the linear approximation (straight black dashed line). $P_0(x)$ is the probability distribution of the unperturbed logistic map. For $i \ge 1$ the probability distribution is influenced by the dynamic of the node in the previous layer, thus $P_i(x)\neq P_0(x)$.}
    \label{fig:probdistr}
\end{figure}

Below we consider a chain of logistic maps at the Ulam point  defined by the recurrence equation 
\begin{equation}
    x'=4x(1-x).
    \label{eq:logisticeq}
\end{equation} 
The stationary probability distribution of the state variable $x$ is $P_0(x) = \left(\pi \sqrt{x(1-x))}\right)^{-1}$, see Fig~\ref{fig:probdistr}. The corresponding Lyapunov exponent is
\begin{equation}\label{eq:logisticdistr}
    \lambda_0 = \int_0^1 \frac{\ln(4|1-2x|)}{p \sqrt{x(1-x)}} \mathrm{d} x=\ln(2).
\end{equation}
This yields, according to the general relation \eqref{eq:pic}, $p^{(c)}=\frac{1}{2}$.

\subsection{Patterns of synchrony in a small chain}

\begin{figure}[!htb]
\centering
   (a)\includegraphics[width=\textwidth]{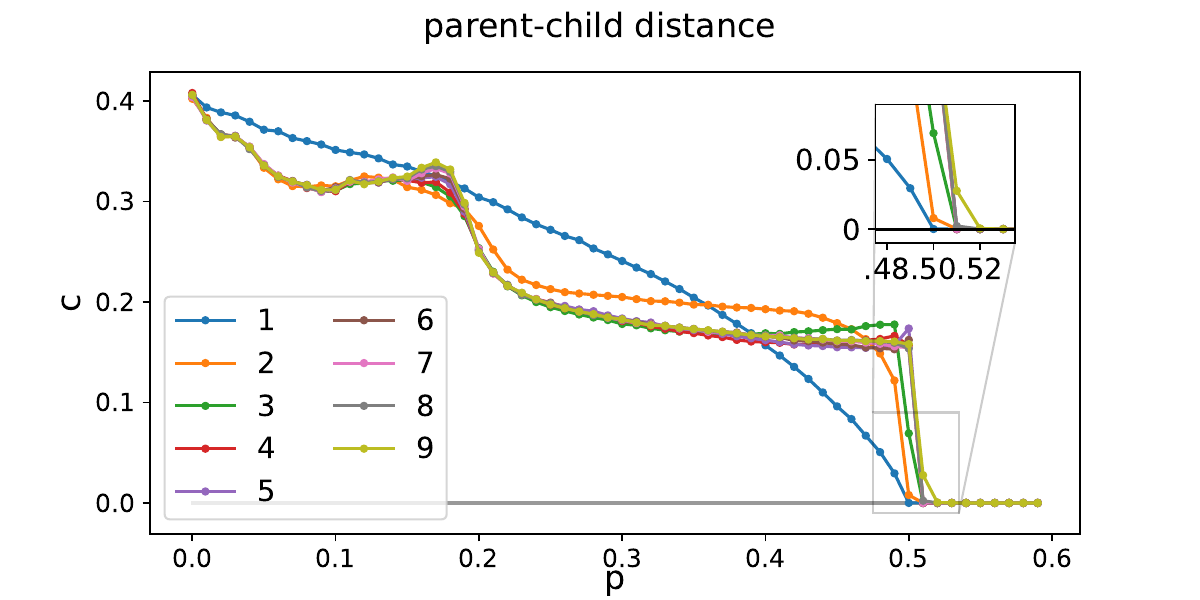}  \\
     (b)\includegraphics[width=\textwidth]{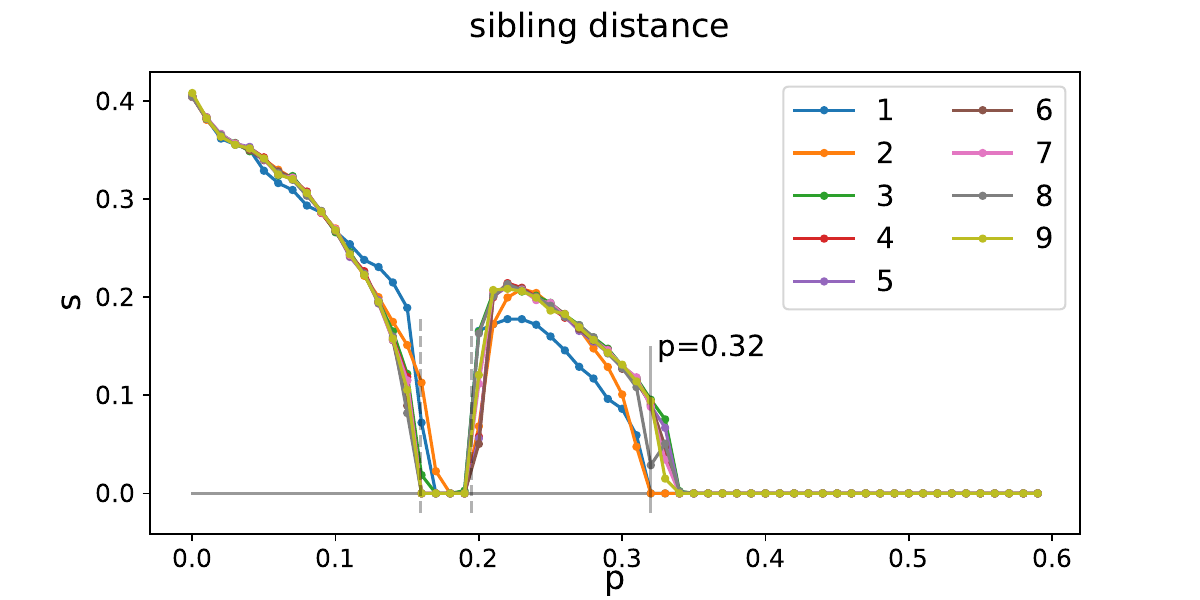}
    \caption{The distances between parent and child  (a) and between the two siblings (b) for the first $10$ layers of the logistic chain; $T=10^4$, $\tau=10^5$.}
    \label{fig:distance_10}
\end{figure}

In the absence of the complete synchronization across the layers, i.e., for $p<p^{(c)}$, the regimes at $i>0$ are different from that of Eq.~\eqref{eq:logisticeq} and the form of the probability density varies with the layer number $i$. Fig.~\ref{fig:probdistr} indicates that for large $i$, the distributions for different $i$ become nearly equal and are well approximated by $P(x)=2x$. Inserting this approximation in the relation \eqref{eq:lyapprob}, we get $\lambda_i=\ln(4)-1\approx 0.386$. The corresponding approximate value of the critical coupling according to Eq.~\ref{eq:pis} is $p_i^{(s)}\approx 0.32$ in good agreement with Fig.~\ref{fig:distance_10}(b).

\begin{figure}[!htb]
    \centering
    \includegraphics[width=\textwidth]{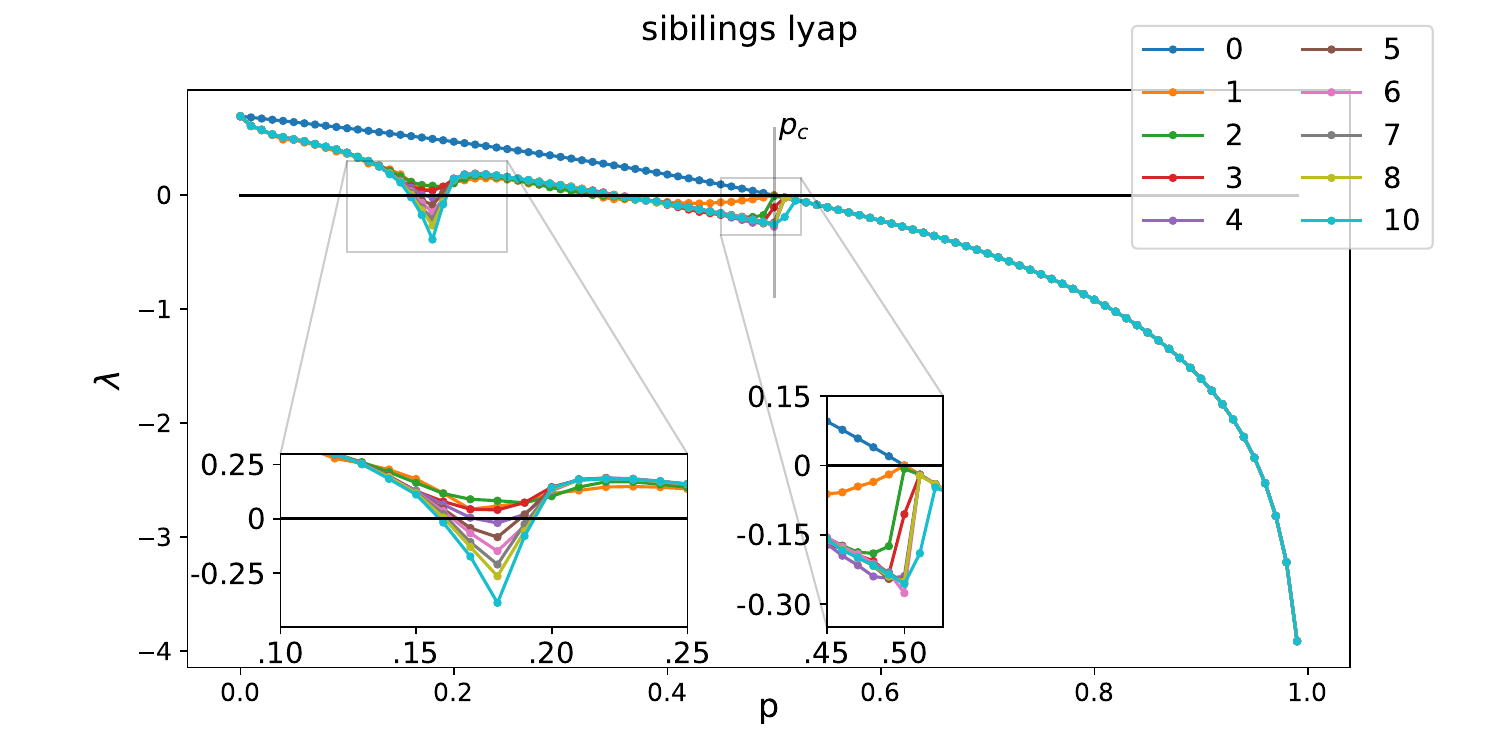}
    \caption{The computed Lyapunov exponent $\lambda_1^{(c)}$ ($0$ in legend) and the Lyapunov exponent $\lambda_i^{(s)}$ of the chain; $T=10^4$, $\tau=10^5$. } 
    \label{fig:lyap_10}
\end{figure}

In Fig.~\ref{fig:distance_10} and Fig.~\ref{fig:lyap_10} we show the results of our simulations with a limited number of nodes with logistic maps:  we plot the average distances $c_i$ (Fig.~\ref{fig:distance_10}(a)) and $s_i$ (Fig.~\ref{fig:distance_10}(b)) for the first 10 layers of the system. In Fig.~\ref{fig:lyap_10} we show the computed sibling Lyapunov exponents, $\lambda_i^{(s)}$, and we also plot the first computed parent-child Lyapunov exponent $\lambda_1^{(c)}$ (labeled $0$ in legend). Note that for $p>p^{(c)}$ all nodes are synchronized, so the Lyapunov exponent follows the same behavior for all nodes.

We can notice that the transition to parent-children synchronization across the layers occurs for all layers around the estimated value (Eq.~\ref{eq:pic})  $p^{(c)}\simeq 0.5$, although with quite a different behavior near the transition threshold.  Synchronization within a layer starts at a value of $p_i^{(s)} \simeq 0.32 < p^{(c)}$.

\subsection{Synchronization window with a chessboard pattern}

\begin{figure}[!htb]
    \centering
    \includegraphics[width=\textwidth]{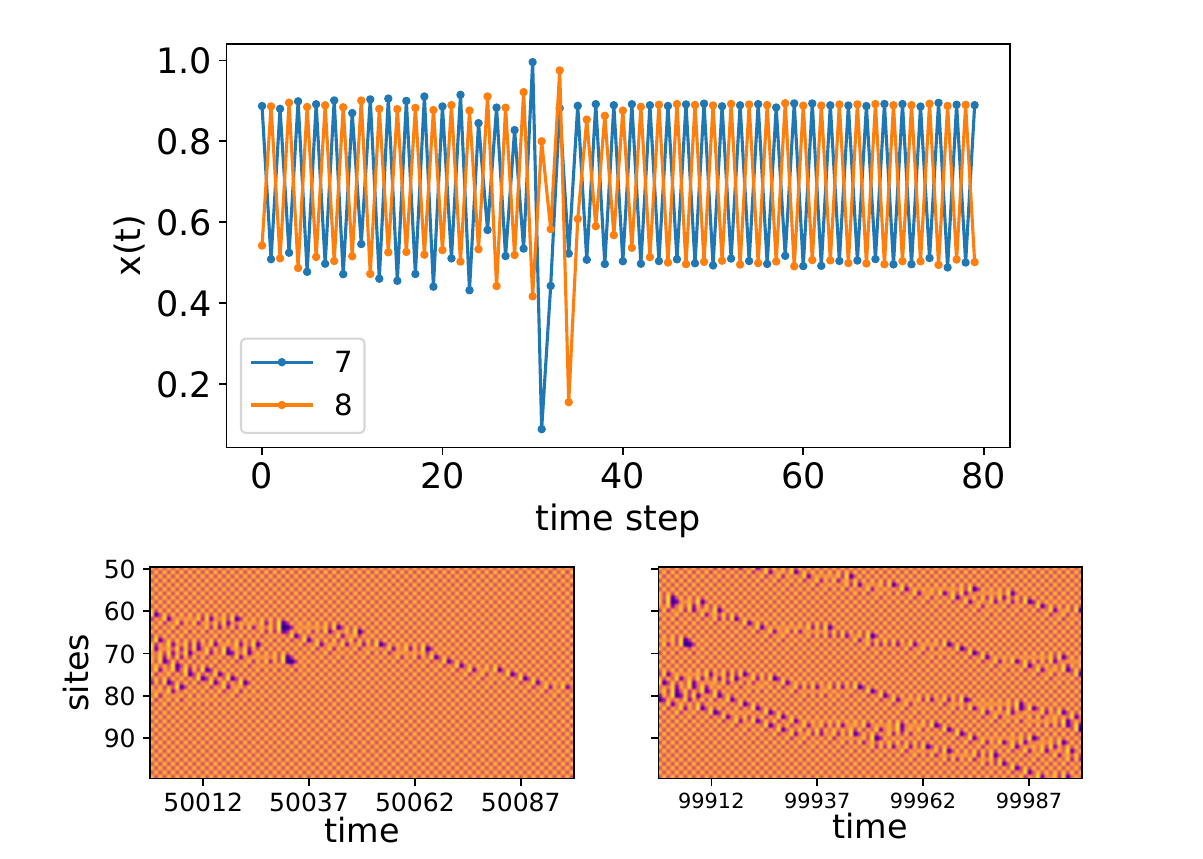}
    \caption{ (Top) Snapshot of the time evolution of maps $x_5(t)$ and $x_6(t)$ for $p=0.18$ and (bottom) heat map of the time evolution of the nodes in the deeper layers.
    In both cases we show the time evolution for each time step. $T=10^{5}$, $\tau=10^4$. 
    }
    \label{fig:p18}   
\end{figure}

There is also a synchronization-within-a-layer window near $p=0.18$, which can be explained as follows. A detailed visual inspection (Fig.~\ref{fig:p18}) reveals that here a chessboard-like spatio-temporal pattern close to having period 2 both in space and time is observed. Let us assume that such a pattern is superstable, i.e. it contains the point $x=1/2$ at which $f'(x)=0$. Such a pattern has a form $x_{2m+1}(2k+1)=x_{2m}(2k)=\frac{1}{2}$, $x_{2m+1}(2k)=x_{2m}(2k+1)=X$ with some unknown $X$. It is a solution of the logistic chain if
\[
\frac{1}{2}=p f\left(\frac{1}{2}\right)+(1-p)f(X),\quad X=pf(X)+(1-p)f\left(\frac{1}{2}\right)\;.
\]
This system of equations has a solution
\[
X=\frac{3+\sqrt{17}}{8},\quad p=\frac{5-\sqrt{17}}{9-\sqrt{17}}\approx 0.1798\;.
\]
One can expect that around this superstable value of the parameter, $0.16\lesssim p\lesssim 0.2$ the pattern will still be stable. However, a perfect period-2 pattern will be only observed for a particular period-2 trajectory at site $i=0$. For a generic chaotic driving $x_0(t)$, one, first, starts to observe this pattern for large enough $i$, and second, this pattern is not perfect. At large distances from edge $i=0$,
one observes long patches of the period-2 behavior intermingled with defects (Fig.~\ref{fig:p18}); the number of defects decreases with $i$. This explains why only at large $i$ the transversal Lyapunov exponent for the stability within a layer becomes negative  (cf. Fig.~\ref{fig:lyap_10}).

There is, however, a visible contradiction between the transversal exponent $\lambda_1^{(c)}$ (Fig.~\ref{fig:lyap_10}) and the observed distance $s_i$ (Fig.~\ref{fig:distance_10}) in the region $p\approx 0.18$. Indeed, even for small $i$ where $\lambda_1^{(c)}>0$, synchronization within a layer is observed.

\begin{figure}[!htb]
    \centering
      (a)  \includegraphics[width=0.7\textwidth]{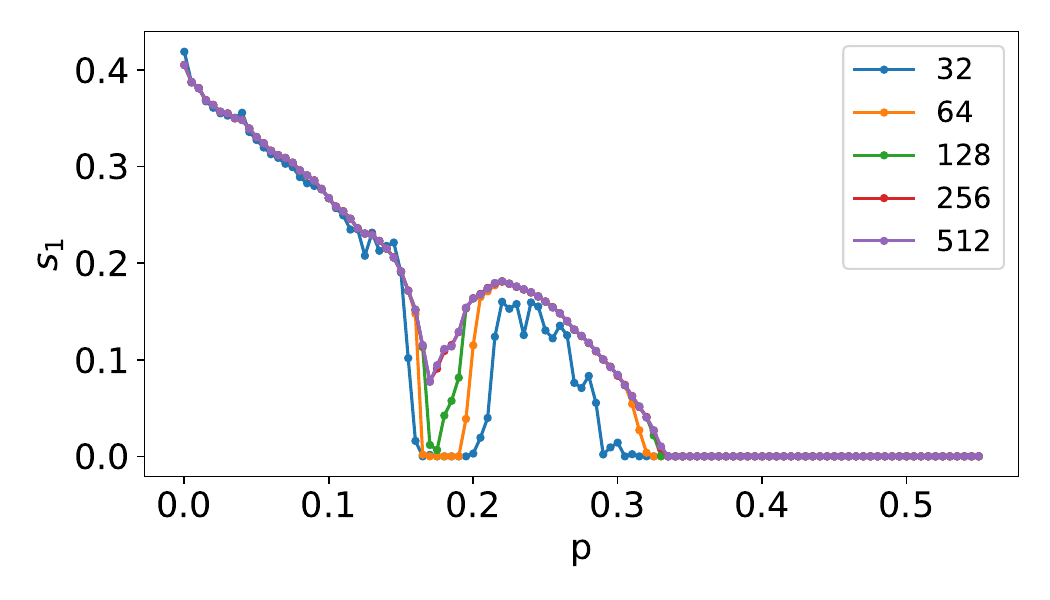} \\
       (b) \includegraphics[width=0.7\textwidth]{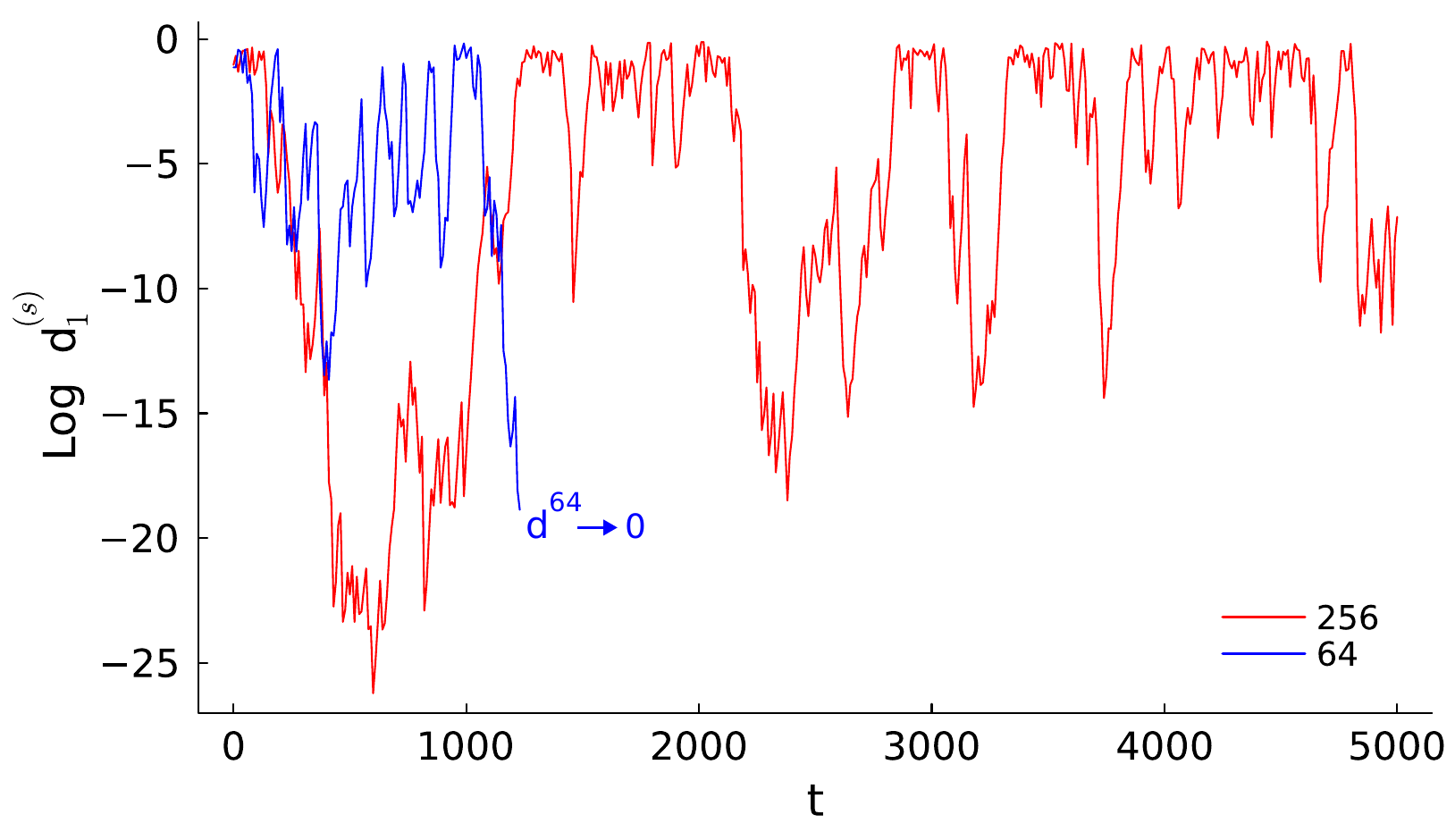} 
    \caption{(a) Siblings distances of the first node ($s_1$) computed with different machine precision $\varepsilon$ (values in legend). Only for precision $ \varepsilon \geq 256$ bit does the synchronization window disappear. Here $T=10^4$, $\tau=10^5$.
    (b) Plot of the time evolution of the siblings distance of the first nodes ($\log_{10}d^{(s)}_1$) computed with $\varepsilon=64$ and $\varepsilon=256$  machine precision at coupling parameter $p=0.18$. The distance with 64-bit precision may eventually reach values lower than machine precision, creating artificial synchronized states.}
    \label{fig:precision}
\end{figure}

This ``anomalous'' synchronization can be understood as a result of finite-precision effects of numerical simulations (cf. Ref.~\cite{Pikovsky-94a}). Even when the transversal Lyapunov exponent is positive, fluctuations in the distance between systems $x_i$ and $y_i$ can reduce this distance to such a small value that the computer representations of the states coincide. Afterward, the states remain identical, despite the presence of transversal instabilities.

To verify this hypothesis, one can rerun the numerical simulations with increased precision $\varepsilon$, for instance, increasing from $\varepsilon=32$ bit up to $\varepsilon=512$ bit, and compare the differences $|x_i-y_i|$ with those obtained in previous simulations using 64-bit precision.

In Fig.~\ref{fig:precision}(a), we present the siblings-distance $s_1=\langle |x_1-y_1| \rangle$, computed at different levels of $\varepsilon$.  As shown, increasing the precision leads to a change in the width of the synchronized window, which vanishes for $\varepsilon > 128$ bit. 

Additionally, in Fig.~\ref{fig:precision}(b), we present an example of the behavior of the sibling distance $d^{(s)}_1(t)=|x_1(t)-y_1(t)|$ at the first node for $p=0.18$. Specifically, we compare this distance at low precision ($\varepsilon=64$) with that at high precision ($\varepsilon=256$). It is evident that higher machine precision leads to more accurate numerical representations, resulting in a minimum distance between the two nodes on the order of $10^{-20}$. At lower precision, synchronization is driven by approximation truncation: at the given coupling parameter, the distance eventually falls below machine precision, becoming numerically zero.

\subsection{Importance of the numerical implementation for large chains}

Before describing the results obtained for the deep system, e.g., with a number of layer~$\gg 10$, let us consider how the coupling scheme can be rewritten in two mathematically equivalent formulations, which, however, yield profoundly different computational results  when combined with the map of Eq.~\eqref{eq:logisticeq}.

Equation~\eqref{eq:map} can be computationally implemented using the \textit{direct coupling} scheme, as defined in the mathematical model:
\begin{equation}
    \begin{split}
    x_i' &= (1-p) f(x_i) + p f(x_{i-1});\\
    y_i' &= (1-p) f(y_i) + p f(x_{i-1}).\\
    \end{split}\label{eq:CSdirect}
\end{equation}
Another possible numerical implementation is a \textit{delta coupling}
\begin{equation}
    \begin{split}
    x_i' &= f(x_i) + p \big(f(x_{i-1}) - f(x_i) \big);\\
    y_i' &= f(y_i) + p \big(f(x_{i-1}) - f(y_i) \big),\\ 
    \end{split}\label{eq:CSdelta}
\end{equation}
where first, a difference of the values at two layers is calculated, and then multiplied by $p$ and added to the value at the current layer.

\begin{figure}[!htb]
    \centering
    \begin{tabular}{cc}
       {\includegraphics[width=0.45\textwidth]{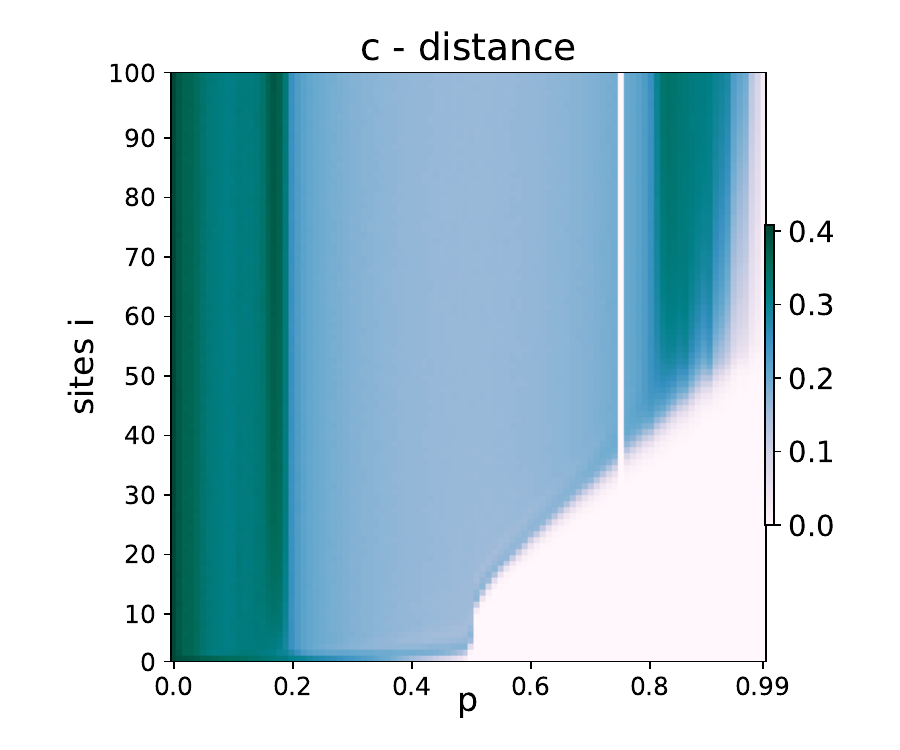}}  & {\includegraphics[width=0.45\textwidth]{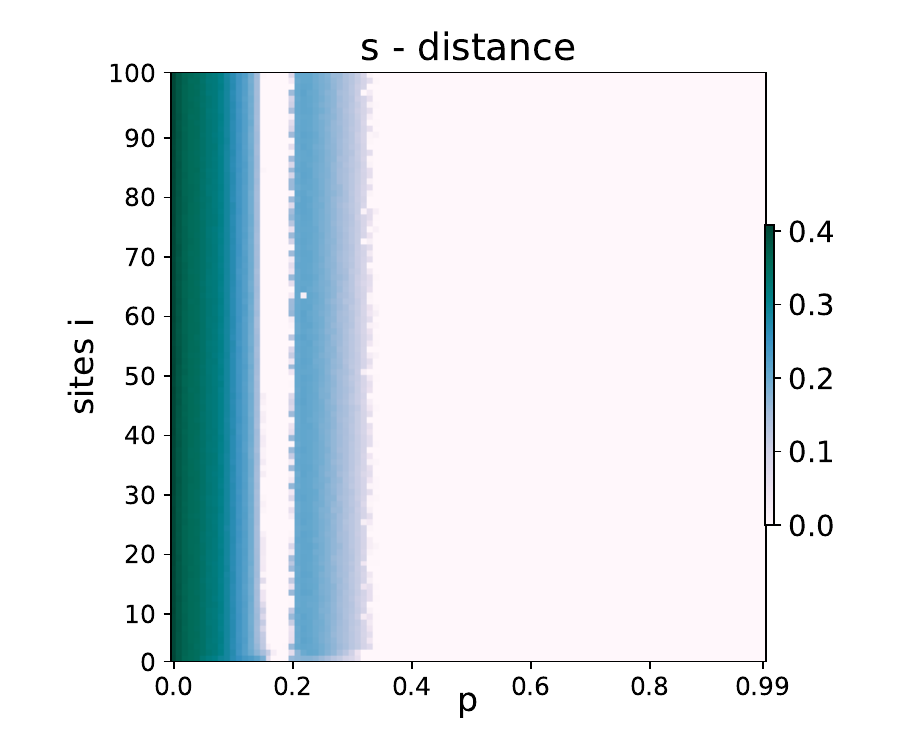}}\\
        (a) & (b) \\
        {\includegraphics[width=0.45\textwidth]{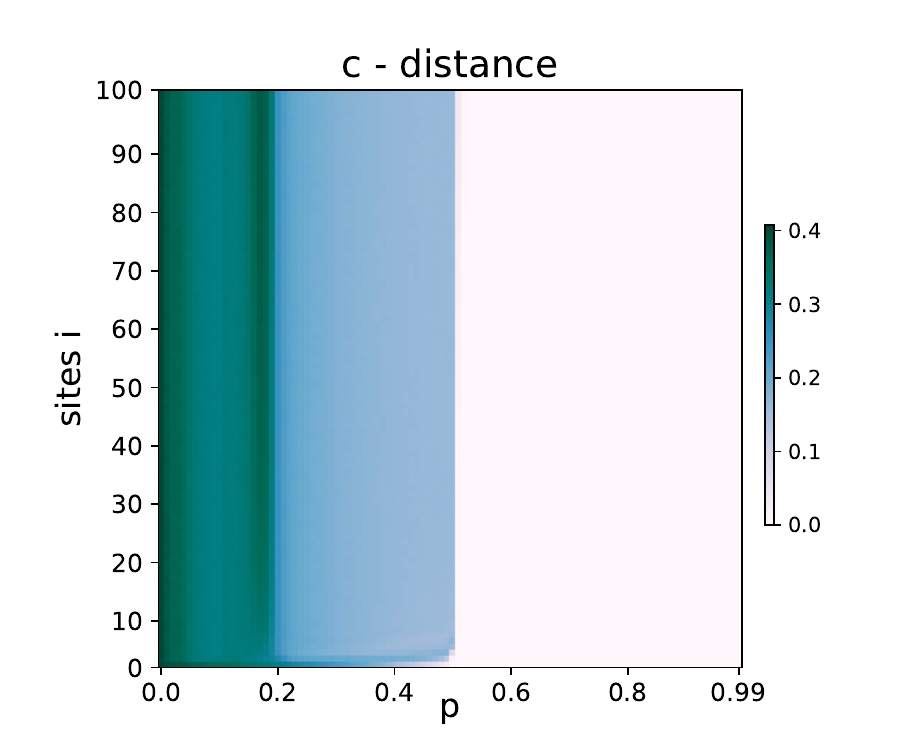}}  & {\includegraphics[width=0.45\textwidth]{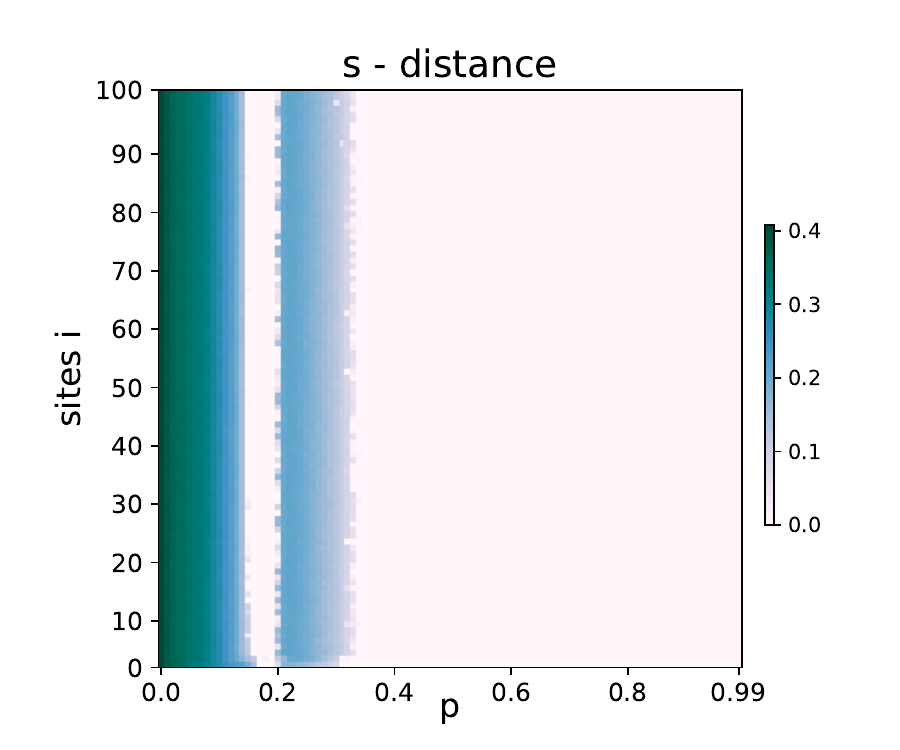}}\\
        (c) & (d) 
    \end{tabular}
    \caption{Heat maps of the parent-child distances (left column) and sibling distances (right column). In the first row we show the results obtained with the \textit{direct coupling} scheme, Eq.~\eqref{eq:CSdirect}, in the second are shown the same plot for the \textit{delta coupling}, Eq.~\eqref{eq:CSdelta}.}
    \label{fig:HeatMapDistances}
\end{figure}    

In Fig.~\ref{fig:HeatMapDistances} we show the results for the distances (parent-child across the layers and siblings within a layer) obtained using the two different computational schemes (Eq.~\eqref{eq:CSdirect}, Eq.~\eqref{eq:CSdelta}). As we can see, the behavior of the synchronization state at $p>p^{(c)}$ in this kind of system is strongly influenced by the computational scheme used to implement Eq.~\eqref{eq:map} for the distance across the layers for deeper nodes ($\displaystyle > 10$). 

One can see that the sibling distance is not affected by the computational schemes used. Instead, the distance across the layers drastically depends on the numerical implementation. In particular, for coupling values greater than the critical value $p^{(c)}$, it is not possible to obtain a completely synchronized state when the simulation is performed with the direct computation scheme.

\subsubsection{Boundary sensitivity of the synchronized dynamics}

The reason for the sensitivity of the dynamics to the numerical scheme lies in the convective instability of the synchronous dynamics. This effect in coupled map lattices has been discussed in Refs.~\cite{aranson1992spatial,Arkady93} in the context of asymmetrically coupled map lattices, which also includes the unidirectional coupled chain of the present study. Indeed, the usual Lyapunov exponent measures the on-site evolution of a perturbation, while a perturbation can decay on-site but propagate along the chain as a pulse with growing amplitude. To identify such a state, one needs to calculate a velocity-dependent (a.k.a. comoving) Lyapunov exponent $\lambda(v)$ which explores growth rates from a localized in space initial perturbation along the spatial-temporal rays propagating with different velocities $v$~\cite{deissler1987velocity,pikovsky2016lyapunov}. Usual Lyapunov exponent appears in this formalism as $\lambda(0)$, and convective instability happens if $\lambda(0)<0$ but $\lambda(v)>0$ for some $v\neq 0$.

The velocity-dependent Lyapunov exponent for an asymmetrically coupled map lattice has been calculated in Ref.~\cite{Arkady93},
in our case 
\begin{equation}
\lambda(v)=\ln 2+(1-v)\ln\frac{1-p}{1-v}+v\ln\frac{p}{v}\;.
\label{eq:vdle}    
\end{equation}
One can easily check that, indeed, for $p>1/2$, the homogeneous synchronous state is convectively unstable.

\subsubsection{Propagation of local disturbance in a convectively unstable regime}
\label{sec:lp}

\begin{figure}[!htb] 
    \centering
    \begin{tabular}{c c}
    \includegraphics[width=0.49\textwidth]{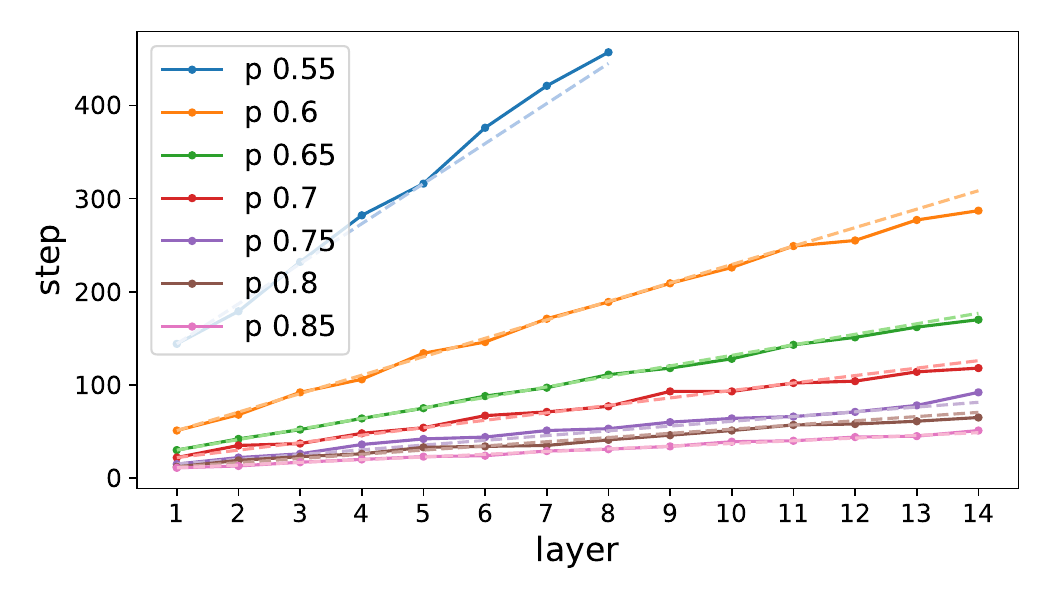}  & \includegraphics[width=0.49\textwidth]{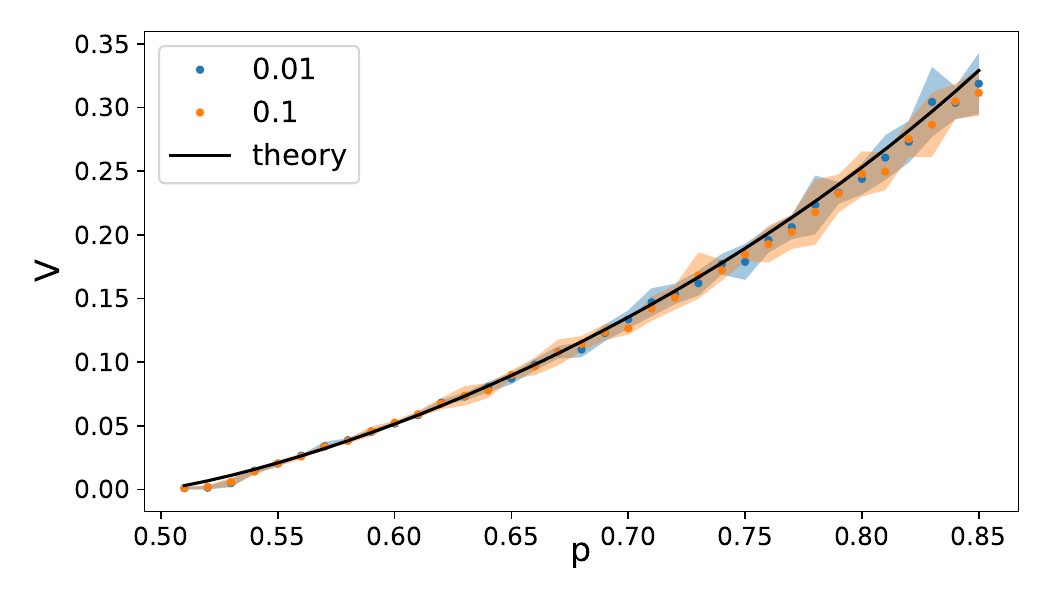} \\
        (a) & (b)
    \end{tabular}
    \caption{(a) Number of time steps needed to reabsorb a perturbation of amplitude $\eta$ from a globally synchronized state. In the dashed line, the linear fit is shown. $T=10^3$, $\eta = 0.001$.
    (b) Velocity versus coupling parameter for different initial noise values $\eta$ (in legend) and, in black line, the theoretical value estimated by solving $\lambda(V)=0$ in Eq.~\eqref{eq:vdle} numerically.}
    \label{fig:logisticNoise_propagation}
\end{figure}

One experiment where this relation can be checked is to calculate the time to re-synchronization if a local perturbation is added.
Assuming the system is in a state of complete synchronization ($p \gtrsim p_c$) we initialize the chain in a synchronized configuration such that $x_i(0)=x_0(0)$. A small perturbation is introduced to the initial condition of the first node $x_0(0)\to x_0(0)+\eta$. This perturbation spreads along the chain, but eventually, all nodes return to the synchronized state. We analyze the time required for each node to resynchronize, utilizing the \textit{delta coupling} scheme for the computational scheme. 
In Fig.~\ref{fig:logisticNoise_propagation}(a), we present, for various coupling values, the number of time steps $t$ needed for the distance $d^{(0)}_i (t) = |x_i(t) - x_0(t)|$ between the first node ($x_0$) and the $i^\mathrm{th}$ node to fall below a predefined threshold (set at $10^{-8}$) when subjected to a disturbance of intensity $\eta = 0.001$. As observed, the behavior is approximately linear. The slope estimated from the data allows us to define a propagation velocity of the perturbation, or more precisely, a velocity at which the disturbance is reabsorbed along the chain. This velocity approaches zero at the critical threshold $p^{(c)}$, indicating the divergence of the reabsorption times.

The reabsorption velocity $V$ can be determined by finding the velocity at which the velocity-dependent Lyapunov exponent (Eq. \eqref{eq:vdle}) vanishes: $\lambda(V) = 0$. In Fig.~\ref{fig:logisticNoise_propagation}(b), we compare the simulation results with theoretical predictions (to find $V$, we found the root of Eq.~\eqref{eq:vdle} numerically), showing that this approximation holds well.

\subsubsection{Spatial amplification of persistent boundary perturbances}
\label{sec:pbp}

\begin{figure}[!thb]
    \centering
    \includegraphics[width=\textwidth]{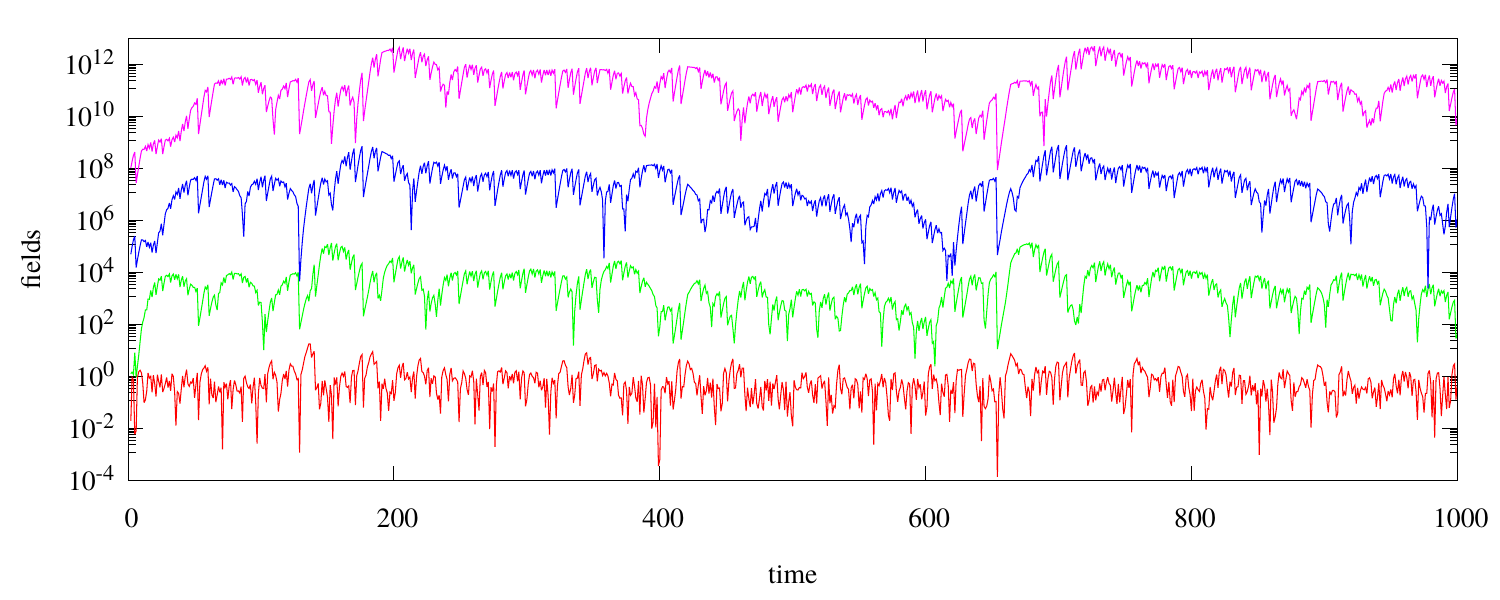}
    \caption{Evolution in space of the perturbation field $|z_n(t)|$ for $p=0.6$. Lines show (from bottom to top) $n=1$ (red), $n=6$ (green), $n=11$ (blue) and $n=16$ (magenta).}
    \label{fig:spgr}   
\end{figure}

In section \ref{sec:lp} we considered a single local perturbation at the boundary, while here we look at a persistent perturbation on top of a synchronous state $x_n(t)=X(t)$ at $p>\frac{1}{2}$. A small perturbation $z_n(t)=x_n(t)-X(t)$ evolves in the linear approximation according to
\begin{equation}
z_n(t+1)=f'(X(t))[p z_{n-1}(t)+(1-p)z_n(t)]\;.
\label{eq:lindyn}    
\end{equation}
Let us consider the \textit{spatial} evolution of the field $z_n(t)$. Namely, we fix the time interval $1\leq t\leq T$ and set a random
homogeneous in time boundary condition $z_0(t)$ 
(e.g., from a uniform distribution). Then we iterate Eq.~\eqref{eq:lindyn} in
space to find the fields $z_1(t), z_2(t),\ldots$. It is convenient, to avoid boundary in time effects, to use periodic boundary conditions
in $t$ (i.e. we set $z_n(1) \equiv z_n(T + 1)$). The field $z_n(t)$ grows with $n$, as illustrated in Fig.~\ref{fig:spgr}. 

\begin{figure}[!htb]
\centering
\includegraphics[width=0.8\columnwidth]{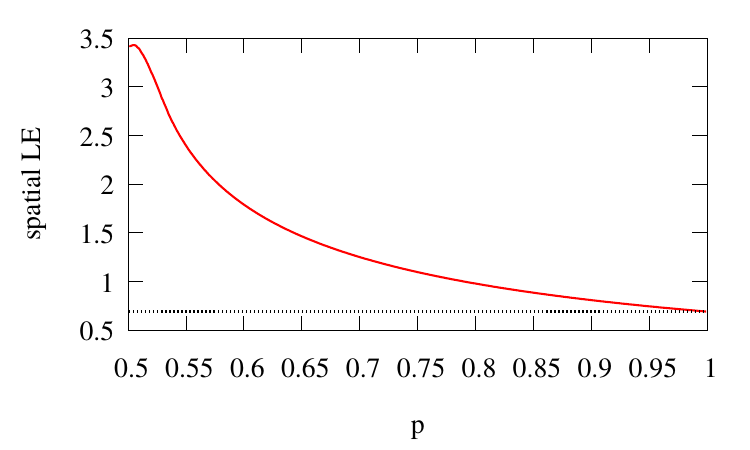}
\caption{Spatial Lyapunov exponent $\Lambda$ (Eq.~\eqref{eq:sle}) in dependence on the coupling parameter $p$. The dotted line is the value $\ln(2)$ to be expected at $p=1$.}
\label{fig:sp2}
\end{figure}

The logarithmic scale in Fig.~\ref{fig:spgr} suggests exponential growth of the perturbations in space.
This can be quantified with the spatial Lyapunov exponent (see \cite{Rudzick-Pikovsky-96} for a growth rate of periodic in time perturbations)
\begin{equation}
\Lambda=\lim_{n\to\infty}\ln \frac{|z_n(t)|}{|z_0(t)|}\;.
\label{eq:sle}    
\end{equation}
This quantity in dependence on $p$ is shown in Fig.~\ref{fig:sp2} for the logistic map. The spatial Lyapunov exponent defines an effective ``boundary layer'' close to the left end $i=0$, within which one can observe synchronization: $\Delta\approx \Lambda^{-1}\ln |v_0|$. One can see that this domain grows with parameter $p$  and is minimal close to the synchronization threshold $p\gtrsim \frac{1}{2}$.

\subsubsection{Numerical scheme as a source of boundary perturbations}

We have observed that the two computational schemes exhibit significantly different results for coupling values greater than the critical value, particularly in the behavior of the deeper layers of the system. This difference arises from a combination of computational effects with the convective instability described above.

Here, we argue that in numerical simulations, persistent boundary perturbations can arise, leading, as has been shown in Section~\ref{sec:pbp}, to the loss of synchrony at large distances from the boundary.

Let us consider the \textit{direct coupling} scheme  (Eq.~\eqref{eq:CSdirect}), and let $t$ be the time instant at which synchronization between two consecutive nodes hold: $x_i = x_{i-1}$. In that case, $f(x_i) = f(x_{i-1})$ also holds; however, due to a finite representation of numbers in a digital computer, it is not automatically true that $(1-p)f(x_i) + pf(x_{i-1}) = f(x_{i-1})$, therefore it can happen that $x_{i}' \neq x'_{i-1}$ (in fact, they differ in computer representation just by one last bit). This can create situations where the states of two consecutive nodes are identical, but the mixing of the factor $p$ and $(1-p)$ desynchronizes them, resulting in $x'_{i} \neq x'_{i-1}$. This, combined with the propagation of perturbations, break the s
tability of synchronized state along the chain.

\begin{figure}[!htb]
    \centering
    \includegraphics[width=0.7\textwidth]{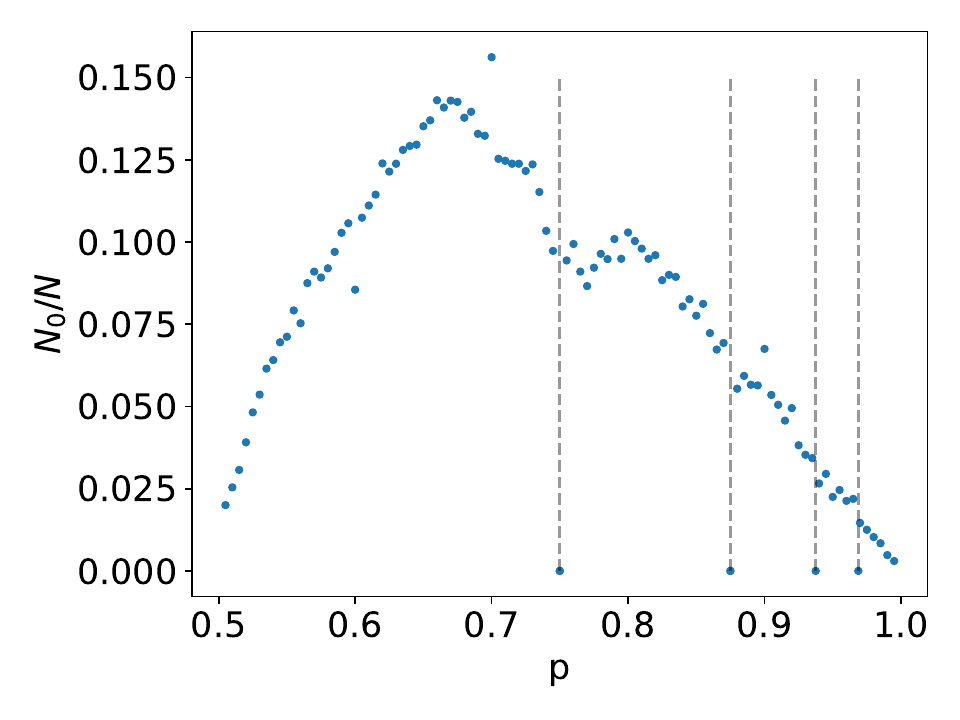}
    \caption{ Fraction of number $N_0$ that can break the synchronization due a mantissa error over $N=10000$ randomly extracted in $[0,1]$ for different values of the coupling parameter $p$. In dashed line we show the values $\Tilde{p} = (2^n - 1 )/ 2^n$.}
    \label{fig:fractionValueDestibilize}   
\end{figure}

The effect of a non-exact computation in the direct scheme is state-dependen, it does not happen for all values of $x_i$.
In Fig.~\ref{fig:fractionValueDestibilize} we plot the fraction of input $x$ from which a perturbation of the synchronized state can arise due to the finite representation of numbers in the computer. Specifically, the plot shows the number $N_0$ of values such that $(1-p)*f(x) + p*f(x) \neq f(x)$ out of $N$ random samples of $x\in [0, 1]$ for various values of $p$. As can be observed, this fraction is non-zero for almost all the considered values of $p$, showing a maximum around $p\approx 0.67$, before decreasing to zero as $p \rightarrow 1$. 

To confirm that this phenomenon is purely artificial and caused by the numerical representation used in the simulation, we note that for certain values (marked by a gray dashed vertical line in figure), the fraction $N_0/N = 0$. These values are $\Tilde{p} = (2^n - 1 )/ 2^n$ for integer $n>2$.
If we examine the binary representation of these numbers, the effect of multiplying $p'=(1-\Tilde{p})$ by $y=f(x)$ is to shift the most significant bit of $y$ by $n$ positions to the right. Since the action of $f$ on $x$ can at most change the first digit from zero to one, the combination of these two effects automatically leads, if $n>2$, to a fictitious synchronized state.
This explains why the complete synchronized state observed in the case of \textit{direct coupling} for $p=0.75$ in Figure~\ref{fig:HeatMapDistances}(a).

Note that the chaotic nature of the local dynamic is crucial: it guarantees that there exists a non-zeros probability such that the value of the state $x_i$ is one of the $N_0$ values that can desynchronize the chain.

In contrast, in the \textit{delta coupling} scheme, this phenomenon does not occur because if $f(x_i) = f(x_{i-1})$, then $p*\big(f(x_{i-1}) - f(x_{i})\big) = 0$ exactly, and not due to approximations resulting from mantissa error.

\section{Conclusions}
In summary, in our study we explored the synchronization dynamics of a branching chain of unidirectionally coupled chaotic maps, focusing on parent-child and siblings synchronization. This configuration captures the essence of process on a Cayley tree structure.

We found that both transitions are governed by the corresponding transversal Lyapunov exponents, which can be calculated according to the probability distribution of the driving unit.  In particular, parent-child synchronization across the layers is governed by the probability distribution of the original chaotic map and the resulting Lyapunov exponent is node-independent. Instead,  sibling synchronization within a layer , which occurs before parent-child synchronization, depends on the probability distribution of the state of parent node, and this vary with the layer. We have shown, however, that a linear distribution provides a good approximation.  

We also analyzed the importance of numerical precision in this type of simulations, showing how synchronization windows can depend on machine precision.
We introduced two possible computational schemes to investigate this problem, observing that, although mathematically equivalent, the two schemes present different behaviors of synchronization across the layers for the deep nodes of the system and large values of the coupling parameter.
In fact, from the simulations, we have observed that with the \textit{direct coupling} scheme (Eq.~\ref{eq:CSdirect}) the synchronization of the deeper nodes is destroyed due to a combination of spurious computational effects enhanced by the chaotic local dynamics and amplified along the chain. 

The convective character of an instability on the synchronous state results in the propagation of a  localized perturbation with re-synchronization. We have checked that the re-synchronization time is well predicted by a velocity-dependent Lyapunov exponent (Eq.~\ref{eq:vdle}). For a persistent irregular perturbation, as it appears, e.g., in a direct numerical scheme, the proper characterization is based on the introduced spatial Lyapunov exponent (Eq.~\ref{eq:sle}). This quantity defines the boundary synchronization region and explains why, at large distances from the end, no synchrony is observed.

Remarkably, the two discussed sources of numerical imprecision play different roles: in one case, it leads to a synchronous state although being unstable; in the other case, it breaks synchrony in spite of his stability. 

We expect that similar results hold for other maps as well, and also in continuous-time systems, like  unidirectionally coupled chains of ordinary differential equations.\\

\small{Author Contributions.
\textbf{Michele Baia:} software, validation, investigation, writing---original draft, writing---review and editing, visualization. \textbf{Franco Bagnoli:} conceptualization, methodology, validation, formal analysis, investigation, writing---original draft, writing---review and editing, supervision. \textbf{Tommaso Matteuzzi:} validation, review and editing. \textbf{Arkady Pikowsky:} methodology, validation, formal analysis, investigation,  writing---review and editing, visualization, supervision.\\
All authors have read and agreed to the published version of the~manuscript.\\

This research did not receive any specific grant from funding agencies in the public, commercial, or not-for-profit sectors.}
    
\bibliographystyle{elsarticle-num} 
\bibliography{synchro}
 
\end{document}